\documentstyle[aps]{revtex}

\begin{document}


\draft

\title{Symmetry-seeking spacetime coordinates}

\author{David Garfinkle} \address{Dept. of Physics, Oakland
University, Rochester, MI 48309} \author{Carsten Gundlach}
\address{Enrico Fermi Institute, University of Chicago, 5640 S. Ellis
Ave., Chicago, IL 60637}

\date{5 August 1999}

\maketitle


\begin{abstract}

In numerically constructing a spacetime that has an approximate
timelike Killing vector, it is useful to choose spacetime coordinates
adapted to the symmetry, so that the metric and matter variables vary
only slowly with time in these coordinates. In particular, this is a
crucial issue in numerically calculating a binary black hole
inspiral. An approximate {\it homothetic} vector plays a role
in critical gravitational collapse.  We summarize old and new
suggestions for finding such coordinates from a general point of view.
We then test some of these in various toy models with spherical
symmetry, including critical fluid collapse and critical scalar field
collapse.

\end{abstract}


\section{Introduction}


One of the fundamental problems in numerically constructing spacetimes
without symmetries (3D numerical relativity) is the choice of suitable
coordinates. The task is not just to find a good coordinate system for
a known spacetime, but the spacetime is not actually known when the
calculation begins, so that one has to construct the coordinate system
along with the spacetime. In principle, once one has obtained the
spacetime in some coordinate system, one can go back over it and paint
on a better coordinate system. In practice, the initial attempt must
not be too bad, or the coordinate system will break down altogether
before one has evolved very far.

One begins the numerical construction of a spacetime with Cauchy data
(three-metric $g_{ab}$, extrinsic curvature $K_{ab}$, and perhaps
suitable matter data) obeying the Hamiltonian and momentum
constraints, in hand. One then has to make a choice of the lapse
$\alpha$ and shift $\beta^a$ based on this information in order to
start constructing the spacetime along with the coordinate system. For
the purpose of this paper, a coordinate condition is a prescription
that maps Cauchy data on a Cauchy surface to lapse and shift fields on
that Cauchy surface. One then uses the same prescription again on each
time slice. We call such coordinate conditions local in
time. Typically they are elliptic (in space), and one specifies
suitable boundary conditions at the outer (large radius) and inner
(black hole excision) boundaries of the numerical domain.

In a generalization of local-in-time coordinate conditions, one may
use a prescription for the first or second time derivatives of the
lapse and/or shift, obtaining a heat equation or wave equation-like
prescription. Such evolution equations for the lapse and shift require
much less computational work than solving an elliptic equation at each
time step. Coordinate conditions for the purpose of our paper are then 
elliptic, parabolic or wave equations for the lapse and shift in terms
of given Cauchy data on a slice.

What are the characteristics of good coordinate conditions? Here we
propose the following desirable criterion: If the spacetime is in fact
stationary, the coordinate condition should make the metric
coefficients explicitly time-independent, independently of how the
initial slice is embedded in the spacetime. If there is an approximate
timelike Killing vector, the metric should be approximately
time-independent. If there is no Killing vector, the natural extension
of our proposal is to minimize some measure of the rate of
change of the metric with time. 

Consider the inspiral phase of a binary system. There is an
approximate timelike Killing vector here, in the sense that all metric
coefficients can be made to evolve on the inspiral time scale instead
of the much shorter orbital time scale. Clearly, this defines a
corotating coordinate system. If the spacetime has an approximate
periodicity (discrete isometry connecting timelike related points), as
in the case of a highly elliptic inspiral, the metric should become
approximately periodic in time through our coordinate condition.

A coordinate system adapted to the presence of a timelike Killing
vector, as we have just described it, is far from unique. It is easy
to see that the family of such coordinate systems is equivalent to the
family of smooth spacelike slices containing one arbitrary but fixed
spacetime point, together with spatial coordinates on the slice: the
coordinate system on the spacetime is obtained by Lie-dragging the
slice and its coordinates along the Killing vector. 

One may know that the final state of an evolution (for example a binary
black hole system) is a Kerr solution, but this does not mean that the
metric coefficients approach the Kerr metric in any of the standard
coordinate systems even as they become asymptotically
time-independent. We require only that our coordinate condition
evolves any given initial slice along the timelike Killing vector, not
that it ``straightens out'' the slice. A ``straightest'' slice in this
spacetime could be defined as one that minimizes for example $\int
{^{(3)}R}^2 dV$, but we shall not try to minimize this at the same
time.

A mathematically similar problem to that of finding good coordinates
in numerical relativity arises in the theory of critical phenomena in
gravitational collapse. The entire rich phenomenology of critical
collapse can be explained and quantitatively predicted using ideas
taken from renormalisation group theory and the theory of dynamical
systems \cite{Gundlach_critreview}. Implicit in this theoretical
framework is the concept of a coordinate prescription (local in time)
such that the metric is ``time''-independent if the Cauchy data
generate a spacetime that is homothetic, or continuously self-similar
(CSS). We have set ``time'' in quotation marks here, as
$\partial/\partial t$ is not everywhere timelike in a homothetic
spacetime. In this context we are only interested in a prescription
that is local in time (in the sense of not containing time derivatives
of the lapse and shift). Such a prescription turns general relativity
into a dynamical system, with CSS spacetimes corresponding to fixed
points.  If the spacetime is discretely self-similar (DSS), we require
our prescription to make the metric periodic in ``time'', with a DSS
spacetime corresponding to a limit cycle.  (The use of maximal slicing
in critical gravitational collapse\cite{kenandme} does part of the job
of turning general relativity into the appropriate sort of dynamical
system.  However, one is still left with the question of what spatial
coordinates to use).

Again, such a coordinate system is far from unique. In the
neighborhood of any data set generating a CSS or DSS spacetime one can
define such a coordinate prescription perturbatively. This
perturbative definition has been enough in order to calculate critical
exponents to several significant digits. A coordinate condition
applicable to all of superspace, with the property that the fixed
points of the resulting dynamical system are precisely CSS spacetimes
and the limit cycles the DSS spacetimes, would define a
renormalisation group for general relativity. Studying its properties
would help us in understanding the general dynamics of general
relativity and issues relating to cosmic censorship.

In this paper, we study the two related cases in a common
framework. In section II, we discuss a number of candidate coordinate
conditions in the absence of symmetries (other than stationarity or
homothety). Our basic strategy is to write down the condition that the
metric is (conformally) time-independent in a 3+1 split, and then
choose four equations out of that set of equations to solve for the
lapse and shift or their time derivatives. This will guarantee that
our criterion is obeyed. If a given prescription obeying this
criterion is otherwise a good coordinate condition must be determined
empirically. In section III, we specialize the general proposals to
spherical symmetry, and we discuss additional proposals that are
specific to spherical symmetry. In section IV, we test some of these
proposals by evolving a slice in a given spherically symmetric
spacetime. (This procedure is simpler and more stable than evolving
the Cauchy data themselves.) We use several spacetimes to test the
stationary case, the CSS case, and the DSS case.


\section{Coordinate choices}


\subsection{3+1 split of the Killing and homothetic equation}


We begin with a homothetic Killing vector $X^a$.  That is
\begin{equation}
\label{homkv}
{\cal L}_X g_{ab} =
{\nabla _a} {X_b}  +  {\nabla _b} {X_a} = 2  \sigma  {g_{ab}},
\end{equation}
where $\sigma $ is a constant. For $\sigma=0$, $X^a$ is a Killing
vector. For $\sigma \ne 0$ a constant we can, without loss of
generality, choose $\sigma = - 1 $ by rescaling $X^a$.  In the
following $\sigma$ only takes the values $0$ and $-1$.  We perform the
usual 3+1 split of the Einstein equation with $X^a$ as the evolution
vector field.  That is, spacetime is foliated by surfaces $\Sigma (t)$
with ${X^a} = {{(\partial /\partial t)}^a}$.  Each surface has a unit
normal vector $n^a$, intrinsic metric ${h_{ab}}= {g_{ab}}+{n_a}{n_b}$
with associated derivative operator $D_a$ and extrinsic curvature
${K_{ab}}= - (1/2) {{\cal L}_n} {h_{ab}}$.  (There is no standard
convention for the sign of $K_{ab}$. Here we use the sign prevalent in
numerical relativity.)  We decompose $X^a$ as $\alpha {n^a} + {\beta
^a} $ with ${\beta ^a} {n_a} = 0$.  We want to know how the quantities
($\alpha, {\beta ^a}, {h_{ab}}, {K_{ab}}$) evolve with time.  We will
get one answer from the homothetic Killing equation and another answer
from the usual ADM equations.  The condition that these two answers
agree then gives conditions that these quantities must satisfy in a
spacetime with a homothetic Killing vector.\cite{beverly,piotr} .

We begin by deriving evolution equations from the homothetic Killing
equation.  From the decomposition of $X^a$, it follows that ${n_a} = -
\alpha {\nabla _a} t$ and therefore that
\begin{equation}
{{\cal L}_X} {n_a} = {\alpha ^{-1}} ({{\cal L}_X} \alpha ) {n_a}.
\end{equation}
Since ${{\cal L}_X}{n_a}$ is a scalar times $n_a$, it follows 
from equation (\ref{homkv})
that 
\begin{equation}
\label{dndt}
{{\cal L}_X}{n_a} = \sigma {n_a} ,
\end{equation} 
and therefore that
\begin{equation}
\label{dalphadt}
{{\cal L}_X}\alpha = \sigma \alpha.
\end{equation}
Then using $0 = {{\cal L}_X} {X^a} = 0 $ we find
\begin{equation}
{{\cal L}_X} {\beta ^a} = 0 .
\end{equation}
It follows from equations (\ref{homkv}) and (\ref{dndt}) that 
\begin{equation} 
{{\cal L}_X} {h_{ab}} = 2 \sigma {h_{ab}}.
\end{equation}
We then find
\begin{equation}
{{\cal L}_X}{K_{ab}} = - {1 \over 2} {{\cal L}_X}{{\cal L}_n} {h_{ab}}
= - {1 \over 2} \left ( {{\cal L}_n}{{\cal L}_X} {h_{ab}} + {{\cal L}_
{[X,n]}} {h_{ab}} \right ),
\end{equation} 
and therefore 
\begin{equation}
{{\cal L}_X}{K_{ab}} = \sigma {K_{ab}}.
\end{equation}


\subsection{Solving for the lapse and shift locally in time}


We begin by looking at those components of the homothetic Killing
equation that contain time derivatives of the lapse and shift. With
the shorthand ${{\cal L}_X} T = \dot T $ for any tensor $T$, these are
\begin{equation}
\label{gaugetimederiv}
\dot\alpha = \sigma\alpha, \qquad
\dot \beta^a =0.
\end{equation}
Given a good initial guess, we could consider these as evolution
equations for the lapse and shift, but this is unlikely to make for a
stable scheme when the symmetry is only approximate and/or the initial
guess is not perfect. Furthermore, it leaves us with the problem of
finding a good initial value for the lapse and shift. Conversely, if
we have found a lapse and shift on the initial time slice by solving
some elliptic equations, the only reason not to use the same
prescription on each time slice would be numerical effort. [If
numerical effort is a concern, one could in effect evolve the lapse
and shift by (\ref{gaugetimederiv}) for a few time steps, before using
the expensive prescription again.]

We now concentrate on the other components of the homothetic Killing
equations. The ADM evolution equations are
\begin{eqnarray}
\dot h_{ab} & \equiv & D_a\beta_b + D_b\beta_a - 2\alpha K_{ab}, \\
\dot K_{ab} & \equiv & \beta^c D_c K_{ab} + K_{ac} D_b \beta^c  
+ K_{bc} D_a \beta^c - {D_a} {D_b} \alpha + \alpha 
\left ( {^{(3)}}{R_{ab}}  
+  K  {K_{ab}}  -  2  {K_{ac}}  {{K_b}^c}
-  {\tau _{ab}}  -  {1 \over 2}   (\rho - \tau )  
{h_{ab}} \right ),
\end{eqnarray}
for any vector field $X^a$.
Here $\rho \equiv {G_{ab}}{n^a}{n^b}$, ${\tau _{ab}} \equiv
{{h_a}^c}{{h_b}^d}{G_{ab}}$, and ${J_a} \equiv -
{{h_a}^c}{n^b}{G_{cb}}$.  Note that in the ADM evolution equations,
$\alpha $ and $\beta ^a$ can be freely specified, so no new
information is obtained by combining the ADM equations with the
equations (\ref{gaugetimederiv}).  But if we substitute the ADM
equations into the components
\begin{equation}
\label{doth}
{{\dot h}_{ab}} = 2 \sigma
{h_{ab}}
\end{equation}
\begin{equation}
\label{dotK}
{{\dot K}_{ab}} = \sigma {K_{ab}}.
\end{equation}
of the homothetic Killing equation, we obtain two symmetric tensor
equations which link the lapse and shift, and their spatial
derivatives, to the Cauchy data, and their spatial derivatives, all
within a single Cauchy surface. 
 
Our task can now be formulated as follows. Out of these 6+6 equations
we want to obtain 3+1 equations that can be solved for the lapse and
shift, for given generic Cauchy data. The remaining 8 equations remain
as consistency conditions which are obeyed if and only if the Cauchy
data actually evolve into a spacetime with a (conformal) Killing
vector. In other words, given Cauchy data on a slice, we want to have
a prescription to calculate a lapse and shift such that one obtains a
reasonable coordinate system on the spacetime evolved from
``reasonable'' generic data, and coordinates adapted to the symmetry
if there is a timelike Killing vector or homothetic vector.

Clearly, there is an infinite range of possibilities.  In a situation
without symmetries, our task is to make one 3-vector (the equation for
the shift) and one scalar (the equation for the lapse) out of two
symmetric 3-tensors.  It is important to note that the equations for
the lapse and shift will generally be coupled, so that the same
equation for, say, the shift can have a different character when
coupled to a different equation for the lapse. Some or all of the
resulting equations can be purely algebraic. Alternatively, they can
involve first or higher spatial derivatives of the lapse and shift. An
elliptic equation or equations would be particularly appealing. Any
prescription obtained for the general case can be reduced to spherical
symmetry, and we shall do this for testing. Additional prescriptions
that explicitly use the spherical symmetry do not generalize back to
the case without symmetries.

One can obtain a 3-vector of equations from a 3-tensor by taking a
divergence, for example
\begin{equation}
\label{minstrain}
D^a \left(\dot h_{ab} - 2\sigma h_{ab}\right) = D^a \dot h_{ab} = 0.
\end{equation}
This equation can be derived from
varying the action
\begin{equation}
\label{confstrain}
\int d^3 x  \sqrt{h}  h^{ac} h^{bd} (\dot h_{ab} - 2\sigma h_{ab}) 
(\dot h_{cd} - 2\sigma h_{cd})
\end{equation}
with respect to the shift.  As $\sigma$ drops out of
(\ref{minstrain}), it could also have been derived from varying the
action
\begin{equation}
\label{strain}
\int d^3 x \sqrt{h} h^{ac} h^{bd} \dot h_{ab} \dot h_{cd}.
\end{equation}
Condition (\ref{minstrain}) was first suggested as a shift condition by
Smarr and York \cite{SmarrYork} under the name minimal strain
shift. As it is derived from a variational principle, one may hope
that it gives good spatial coordinates also in the absence of an exact
symmetry. Written out explicitly, minimal strain shift is
\begin{equation}
\label{minstshift}
{D_b} {D^b} \beta^a + D_b D^a \beta^b - 2 D_b(\alpha K^{ab}) = 0.
\end{equation}

Taking the divergence of (\ref{dotK}), we obtain the 3-vector equation
\begin{equation}
D^a \left(\dot K_{ab} - \sigma K_{ab}\right) = 0.
\end{equation}
At first sight this is less appealing than (\ref{minstrain}) because
it contains third spatial derivatives of the lapse.

In order to obtain a scalar equation for the lapse, we can take the
trace of (\ref{doth});
\begin{equation}
h^{ab} \left(\dot h_{ab} - 2\sigma h_{ab}\right) = 0
\end{equation}
is equivalent to
\begin{equation}
\dot{\sqrt{h}} - 3\sigma \sqrt{h} = 0,
\end{equation}
that is, it determines the scaling of the 3-volume element. ($h$ is
the determinant of $h_{ab}$.) Written out in terms of $\alpha$ and
$\beta^a$, this equation is
\begin{equation}
\label{trace_doth}
D_a\beta^a - \alpha K - 3\sigma = 0.
\end{equation}
This does not seem to be in general a good equation for the lapse
(since it is ill defined where $K$ vanishes), 
nor (by virtue of being a scalar) for the shift.

The equation resulting from instead contracting (\ref{doth}) with
$K_{ab}$, 
\begin{equation}
\label{confstrainshift}
K^{ab} \left(\dot h_{ab} - 2 \sigma h_{ab}\right) = 0
\end{equation}
can be derived from varying the functional (\ref{confstrain})
with respect to the lapse.  
We shall refer to Eq. (\ref{confstrainshift}) on its own
as the conformal strain lapse for $\sigma=-1$, or the minimal strain
lapse for $\sigma=0$. It is an algebraic equation for the lapse,
\begin{equation}
\label{bctlapse}
\alpha = {K^{ab} D_a\beta_b - \sigma K \over K^{ab} K_{ab}},
\end{equation}
which is linear in the shift. The fact that the lapse may be zero,
negative, or even blow up (where $K_{ab}=0$), is worrisome. Clearly
this prescription cannot be used on all initial data.  For the Killing
case, $\sigma=0$, Brady, Creighton and Thorne (BCT) \cite{BCT} have
recently proposed using the minimal strain lapse and minimal
strain shift together, as both are derived from the same variational
principle (\ref{strain}).  If one substitutes the formal minimal
strain lapse (\ref{bctlapse}) into the minimal strain shift equation
(\ref{minstshift}), it is not clear if the resulting differential
equation for the shift alone is elliptic. We shall refer to the
combination of minimal strain lapse and minimal strain shift, in the
Killing case $\sigma=0$, as the BCT gauge, and for any $\sigma$ as the
generalized BCT gauge.

The equation
\begin{equation}
\label{trace-Kdot}
h^{ab} \left(\dot K_{ab} - \sigma K_{ab}\right) = 0
\end{equation}
is an elliptic equation for the lapse, and seems not to have been
considered before. We shall call it the trace-Kdot lapse. Adding to
this the minimal strain lapse equation, however, we obtain
\begin{equation}
\label{constantK}
0 = h^{ab} \left(\dot K_{ab} - \sigma K_{ab}\right) 
- K^{ab} \left(\dot h_{ab} - 2\sigma h_{ab}\right) 
=  \dot K + \sigma K    .
\end{equation}
For $\sigma=0$, this is known as constant (extrinsic) curvature
slicing.  If $K=0$ initially, and therefore at all times, we have
maximal slicing.  For $\sigma=-1$ we could speak of exponential
curvature slicing. Written out in terms of the Cauchy data, the
equation is
\begin{equation}
\label{genmax}
-D_aD^a\alpha+\left[{}^{(3)}R+K^2 +{1\over2}(\tau-3\rho)\right]\alpha +
\beta^aD_aK + \sigma K = 0.
\end{equation}
Note that we do not assume that $K$ is constant within each slice.
We shall refer to the combination of equations (\ref{minstshift}) and
(\ref{genmax}) as the generalized Smarr-York (SY) gauge.  For an initial
$K=0$ slice, the generalized SY gauge reduces to maximal slicing with
minimal strain.  

The differential equations for lapse and shift must be supplemented
with boundary conditions in order to yield a solution.  For an
approximate timelike Killing field in an asymptotically flat
spacetime, the condition $\alpha \to 1 $ and ${\beta ^a} \to 0 $ at
infinity should be reasonable.  For the case of binary black hole
inspiral, BCT \cite{BCT} have proposed a boundary condition to use
with their equation.  It is less clear what would be a reasonable
boundary condition to use for the case of critical gravitational
collapse.


\section{Spherical symmetry}


\subsection{Reduction to spherical symmetry}


In spherical symmetry, the metric has the form
\begin{equation}
d {s^2} = ( - {\alpha ^2} + {\beta ^r}{\beta _r})d {t^2}  +  
2  {\beta _r}  d r  d t   +   {g_{rr}}  d {r^2}   +   
{g_{\theta \theta }}  ( d {\theta ^2}  +  {\sin ^2} \theta  
d {\phi ^2} ) 
\end{equation}
where $\alpha , {\beta ^r}, {g_{rr}} $ and $g_{\theta \theta }$ are
functions of $r$ and $t$, and where $\beta_r=g_{rr}\beta^r$.
Equations (\ref{doth}) and (\ref{dotK}) now have only two independent
components each.  The $rr$ and $\theta \theta $ components of equation
(\ref{doth}) are respectively
\begin{eqnarray}
\label{hdotrr}
{1 \over 2}   {\beta ^r}   {{d {g_{rr}}} \over {dr}}   +   {g_{rr}}
  {{d {\beta ^r}} \over {dr}}   -   \alpha  {K_{rr}} - \sigma  
{g_{rr}}  &=& 0     , \\
\label{hdotqq}
{1 \over 2}   {\beta ^r}   {{d {g_{\theta \theta }}} \over {dr}}   -  
\alpha  {K_{\theta \theta }} - \sigma  {g_{\theta \theta }} &=& 0      .
\end{eqnarray}
The $rr$ and $\theta \theta $ components of equation (\ref{dotK}) are
respectively 
\begin{eqnarray}
\label{Kdotrr} \nonumber
- {{{d^2} \alpha}\over {d{r^2}}}   +   {1 \over {2
{g_{rr}}}}   {{d {g_{rr}}} \over {dr}}   {{d \alpha} \over {dr}} +
{\beta ^r}   {{d {K_{rr}}} \over {dr}}   +   2 {K_{rr}}   {{d
{\beta ^r}} \over {dr}} && \\
  +   \alpha \left [ {^{(3)}}{R_{rr}}  +  {K_{rr}} \left ( 2  
{{K_{\theta \theta}} \over {g_{\theta \theta }}}   -   {{K_{rr}} \over 
{g_{rr}}} \right ) - {{\tau _{rr}} \over 2}   +   {{{g_{rr}}      
{\tau _{\theta \theta }}} \over {g_{\theta \theta }}}   -   {{\rho
{g_{rr}}} \over 2} \right ]   -   \sigma {K_{rr}} &=& 0, \\
\label{Kdotqq}
- {1 \over 2 g_{rr}}   {{d{g_{\theta \theta }}} \over {dr}}   {{d \alpha} \over
{dr}} + {\beta ^r}    {{d {K_{\theta \theta }}} \over {dr}}   +  
\alpha \left[  {^{(3)}}{R_{\theta \theta }}   +   {{K_{rr}}
{K_{\theta \theta }} \over {g_{rr}}}    +   {1 \over 2} \left(
{\tau_{rr}\over g_{rr}}  -  \rho \right)
{g_{\theta \theta }} \right]  -   \sigma  {K_{\theta
\theta }}
& = & 0.
\end{eqnarray}

The constraint equations
\begin{equation}
\label{sphconstraints}
{^{(3)}}R  +  {K^2}  -  {K_{ab}}{K^{ab}}  -  2 \rho = 0,
\qquad 
{D_a} {{K^a}_b}  -  {D_b} K  -  {J_b} = 0, 
\end{equation}
become in spherical symmetry
\begin{equation}
\label{sphHam}
\rho = {{{^{(3)}}{R_{rr}}}\over{2{g_{rr}}}}   +   {{{^{(3)}}{R_{\theta 
\theta }}} \over {g_{\theta \theta }}}   +   {{K_{\theta \theta}} \over
{g_{\theta \theta }}}   \left ( {{2{K_{rr}}}\over {g_{rr}}}   +   
{{K_{\theta \theta }}\over {g_{\theta \theta }}}\right ),
\end{equation}
\begin{equation}
\label{sphmom}
2   {{d{K_{\theta \theta }}}\over {dr}} = -  {g_{\theta \theta }} {J_r}
   +   \left ( {{K_{rr}}\over {g_{rr}}}   +   {{K_{\theta \theta}}
\over {g_{\theta \theta }}}\right )   {{d {g_{\theta \theta }}}\over {dr}}.
\end{equation}


\subsection{Generic coordinate conditions applied to spherical symmetry}


In spherical symmetry, the conformal strain lapse, 
equation (\ref{bctlapse}) is given by
\begin{equation}
\label{spBCTlapse}
\alpha = {{{{K_{rr}}\over{{({g_{rr}})}^2}}\left ( {1\over 2}{\beta ^r}
{{d{g_{rr}}}\over {dr}}+{g_{rr}}{{d{\beta ^r}}\over {dr}}\right ) +
{{K_{\theta \theta}}\over {{({g_{\theta \theta}})}^2}}{\beta ^r}
{{d{g_{\theta \theta}}}\over{dr}}- \sigma \left ( {{K_{rr}}\over{g_{rr}}}
+ 2 {{K_{\theta \theta}}\over {g_{\theta \theta}}}\right ) }
\over {{{\left ( {{K_{rr}}\over {g_{rr}}}\right ) }^2} +
2 {{\left ( {{K_{\theta \theta}}\over {g_{\theta \theta }}}\right ) }^2}}}.
\end{equation}
The exponential curvature lapse, equation (\ref{constantK}), is given by
\begin{eqnarray}
\label{spconstK}
\nonumber
- {{{d^2}\alpha}\over{d{r^2}}} + \left ( {1 \over {2 {g_{rr}}}}
{{d{g_{rr}}}\over{dr}} - {1\over {g_{\theta\theta}}}
{{d{g_{\theta\theta}}}\over{dr}}\right ) {{d\alpha}\over{dr}} && \\
\nonumber 
+\alpha {g_{rr}}\left [ {{\left ( {{K_{rr}}\over{g_{rr}}}\right )}^2} +
2 {{\left ( {{K_{\theta \theta}}\over{g_{\theta \theta}}}\right )}^2}
+ {1 \over 2}\left ( \rho + {{\tau _{rr}} \over{g_{rr}}} + 2 
{{\tau_{\theta \theta}}\over{g_{\theta \theta}}}\right ) \right ] && \\ 
+ {g_{rr}}\left [ {\beta ^r} {d \over {d r}} \left ( {{K_{rr}}\over {g_{rr}}}
+ 2 {{K_{\theta \theta}}\over {g_{\theta \theta }}}\right ) + \sigma
\left ( {{K_{rr}}\over {g_{rr}}} + 2 {{K_{\theta \theta}}\over 
{g_{\theta \theta}}}\right ) \right ] = 0 .
\end{eqnarray}
The minimal strain shift, equation (\ref{minstshift}), is given by
\begin{eqnarray}
\label{spminst}
\nonumber
{{{d^2}{\beta ^r}}\over{d{r^2}}} + {{d{\beta ^r}}\over{dr}} \left ( 
{1 \over {2{g_{rr}}}} {{d{g_{rr}}}\over{dr}} + {1 \over {g_{\theta \theta}}}
{{d{g_{\theta \theta }}}\over {dr}}\right ) && \\
\nonumber 
+ {{\beta ^r}\over 2} \left [{1 \over {g_{rr}}}
 {{{d^2}{g_{rr}}}\over{d{r^2}}} + 
\left ( {1 \over {g_{\theta \theta }}} {{d{g_{\theta \theta }}}\over {dr}}
- {1 \over {g_{rr}}} {{d{g_{rr}}}\over {dr}}\right ) 
{1 \over {g_{rr}}} {{d{g_{rr}}}\over {dr}}
- {{\left ( {1 \over {g_{\theta \theta }}} {{d {g_{\theta \theta }}}
\over {dr}}\right ) }^2} \right ] && \\
- \left [ {d \over {dr}} \left ( {{\alpha {K_{rr}}}\over {g_{rr}}} \right ) 
+ {\alpha \over {g_{\theta \theta }}} {{d{g_{\theta \theta }}}\over {dr}}
\left ( {{K_{rr}}\over {g_{rr}}} - {{K_{\theta \theta }}\over {g_{\theta 
\theta }}}\right ) \right ] = 0 .
\end{eqnarray}  

Thus, in spherical symmetry the generalized SY gauge is the combination
of equations (\ref{spconstK}) and (\ref{spminst}), while the generalized
BCT gauge is the combination of equations (\ref{spBCTlapse}) and 
(\ref{spminst}).


\subsection{Coordinate conditions tailored to spherical symmetry}


In spherical symmetry, one can also find simpler equations for the
lapse and the shift.  We note that equation (\ref{hdotqq}) contains no
derivatives of the shift, while equation (\ref{Kdotqq}) contains only
one derivative of the lapse.  We can use equation (\ref{hdotqq}) to
express the shift as
\begin{equation}
\label{sph_alg_shift}
\beta^r=2 { \sigma g_{\theta\theta} + \alpha K_{\theta\theta} \over
{d {g_{\theta \theta }}/ {dr}} }.
\end{equation}
For $\sigma=0$ this is called the area freezing shift, as it keeps the
area of a surface for constant $r$ constant. Similarly, we call
(\ref{Kdotqq}) the area lapse. Substituting (\ref{sph_alg_shift}) into
(\ref{Kdotqq}), we obtain a first-order differential equation for
$\alpha$ alone. We can use the momentum and Hamiltonian constraints
(\ref{sphmom}, \ref{sphHam}) to simplify it to the form
\begin{equation}
\label{sph_simple_lapse}
{{d \alpha} \over {dr}} = {{g_{\theta \theta }} \over {d {g_{\theta \theta}}
/dr}}   \left [ \alpha \left ( -  {^{(3)}}{R_{rr}}   -   
{{2{g_{rr}}{K_{\theta \theta }}{J_r}}\over {d {g_{\theta \theta }}/dr}}
  +   {\tau_{rr}}  + \rho {g_{rr}} \right )   +   2 \sigma 
\left ( {K_{rr}}   -   {{{g_{rr}}{g_{\theta \theta }}{J_r}} \over
{d {g_{\theta \theta} }/dr}} \right ) \right ].
\end{equation}
We can integrate this first-order ordinary differential equation from
the center of spherical symmetry outwards.  However, the equation
requires as a boundary condition the value of the lapse at one value
of $r$.  We refer to the combination of equations
(\ref{sph_alg_shift}) and (\ref{sph_simple_lapse}) as the area gauge.


\subsection{Boundary conditions for the lapse and shift equations in
spherical symmetry}


In three space-dimensions without spherical symmetry, some of the
equations for the lapse and shift we have discussed are coupled
elliptic equations. These require boundary values for the lapse and
shift on a surface (topologically a 2-sphere) at large radius. One
would expect the equivalent problem in spherical symmetry to be a
second-order ODE boundary value problem, with one boundary condition
each at $r=0$ and large $r$, for each of $\alpha$ and $\beta^r$. One
boundary condition at $r=0$ is immediately obtained from symmetry,
namely that $\alpha$ must have an expansion in even powers of $r$ and
$\beta^r$ an expansion in odd powers of $r$. For the second boundary
condition, it is as natural to specify $\alpha$ and $\lim
(\beta^r/r)$ at $r=0$ as it would be to specify $\alpha$ and
$\beta^r$ at some large value of $r$. Doing that, we have the great
simplification of integrating an ODE system from $r=0$ outwards,
instead of solving a boundary value problem. 

The equations for $\alpha$ and $\beta^r$ are homogeneously linear in
$\alpha$, $\beta^r$ and $\sigma$. The boundary values we need can be
thought of as an overall factor in both $\alpha$ and $\beta^r/r$,
and their ratio. The ratio is uniquely determined at the center of
spherical symmetry. In general we have
\begin{equation}
-\alpha K + D_a \beta^a = {\partial(\ln\sqrt{g})\over \partial t} = 3
\sigma,
\end{equation}
where the first equality follows from the definition of $K$, while the
second equality follows from (\ref{doth}).  At the center of spherical
symmetry, this yields
\begin{equation}
\beta^r = \left(\sigma + {1\over3}K\alpha\right)r+O(r^3).
\end{equation}
The same result is also obtained by evaluating either
(\ref{trace_doth}), (\ref{bctlapse}) or (\ref{sph_alg_shift}) at
$r=0$.

It remains to determine the overall factor in both $\alpha$ and
$\beta^r$. From (\ref{dalphadt}) we have that $\alpha=\alpha_0={\rm
const.}$ at the origin in the Killing case. The value of $\alpha_0$ is
a matter of convention that only affects the norm of the Killing
vector field, and we can set it to any value that is constant in $t$.

In the self-similar case, we formally have $\alpha=\alpha_0\exp\sigma
t$ at the center, but this result is not very useful, as only one
choice of $\alpha_0$ gives rises to a compatible coordinate system. To
see this, label the central geodesic by proper time $T$ with the
origin chosen so that $T=0$ is the singular point of the self-similar
spacetime. In the self-similar time coordinate $t$ this point is
$t=\infty$. Let $T=T_0$ where the initial slice $t=0$ intersects the
central geodesic. Then it is easy to see that the proper relation
$T=T_0 \exp\sigma t$ is obtained only if we choose $\alpha_0=\sigma
T_0$. In other words, we must guess correctly how far the slice is
away from the singular point.

A prescription for $\alpha$ at the center can be obtained by noting
that we have ${\cal L}_X T_{ab}=0$ and therefore ${\cal
L}_X\rho=-2\sigma\rho$. In the 3+1 split this is
\begin{equation}
(\alpha n^a+\beta^a)\nabla_a\rho = -2\sigma\rho.
\end{equation}
We combine this with the equation for $n^a\nabla_a\rho$ obtained from
stress-energy conservation,
\begin{equation}
0=-n_a\nabla_b G^{ab}=-K\rho + n^a\nabla_a\rho + D_aJ^a
-\tau^{ab}K_{ab} + 2 J^aD_a\ln\alpha.
\end{equation}
At the center of spherical symmetry, $D_a\alpha$ and
$\beta^a$ both vanish, and we obtain an algebraic equation for $\alpha$
alone. Furthermore at the center we have
$K_{ab}=(K/3)g_{ab}$ and a similar expression for $\tau_{ab}$, which
simplifies our final result to
\begin{equation}
\label{CSSalphacenter}
\alpha = {2 \sigma \rho g_{rr} \over 3\left[{dJ_r\over dr} - K_{rr}
\left(\rho + {\tau_{rr}\over g_{rr}}\right)\right]} + O(r^2)
\end{equation}
This equation gives a (nonzero) value for $\alpha$ at the origin 
provided that neither the numerator nor the denominator vanishes there.


\section{Tests in spherical symmetry}


\subsection{Numerical method}


We now discuss empirical tests of some of the coordinate conditions we
have discussed for some given spacetimes, restricting ourselves to
spherically symmetric examples. Rather than to evolve Cauchy data via
the ADM equations, we evolve an embedded slice in the spacetime. All
indices $A,B,C,i,j,k$ in this section denote coordinate components of
tensors, rather than abstract tensor indices. Let a spacetime metric
be given in closed form in coordinates $x^A$, giving rise to a metric
$g_{AB}$. A three-dimensional spacelike slice embedded in the
spacetime, together with a choice of three coordinates $x^i$ on the
slice, is parameterized by four functions of three coordinates,
$x^A=F^A(x^i)$. The four-metric $g_{AB}$ induces a three metric
$g_{ij}$ and extrinsic curvature $K_{ij}$, and there is a unit
timelike normal $n^A$ to the slice at each point.

In our code, we choose a spacetime, and an initial slice, and
calculate $g_{ij}$ and $K_{ij}$. The equations for doing this are derived
in full in \cite{exactpaper}. If the spacetime contains matter, the
components $\rho$, $J_i$ and $S_{ij}$ of the stress-energy tensor are
also computed from $T_{AB}$. These are handed to a 
coordinate condition solver that returns $\alpha$ and $\beta^i$. The
slice is then evolved via
\begin{equation}
{\partial F^A\over \partial t} = \alpha n^A + \beta^i {\partial
F^A\over\partial x^i}.
\end{equation}
We finite-difference the equations using a type of iterative
Crank-Nicholson (ICN) algorithm. $\partial F^A/\partial t$ is computed at
time $t_n$, and the $F^A$ are evolved to $t_{n+1}$ with a forward in
time, centered in space (predictor) step. $\partial F^A/\partial t$ is
then computed at time $t_{n+1}$, and the average of $\partial
F^A/\partial t$ at $t_n$ and $t_{n+1}$ is then used for a corrector
step. The corrector step is iterated to convergence.  To maintain
stability, we use numerical viscosity.

In spherical symmetry, there are only two functions $X^0(r)$ and
$X^1(r)$, and two components each of the induced metric and extrinsic
curvature, $g_{rr}(r)$, $g_{\theta\theta}(r)$, $K_{rr}(r)$, and
$K_{\theta\theta}(r)$, and stress-energy components $\rho(r)$,
$J_r(r)$ and $\tau_{rr}(r)$.  There is only one shift component,
$\beta^r(r)$, and the lapse, $\alpha(r)$. The coordinate condition
solver and slice evolution code are independent and communicate only
through the fields intrinsic to the slice, so that the coordinate
condition solver never sees the $F^A$ directly. The code assumes that
surfaces of constant $t$ are spacelike, and complains if, through a
bad choice of coordinate, they are not, but no assumptions are made on
the coordinates $X^0$ and $X^1$. They could, for example, be null
coordinates.

Integrating the embedding of a slice, rather than Cauchy data, has two
advantages. The first one is stability: empirically, the ICN
algorithm, together with a linear extrapolation outer boundary
condition, is very stable for given $\alpha$ and $\beta^r$. In
particular, if we want to excise the center of spherical symmetry from
the numerical domain in a black hole spacetime, an extrapolation inner
boundary works well in the slice evolution code, while black hole
excision in a genuine spacetime evolution code is a nontrivial problem
even in spherical symmetry.  The second advantage of the slice
evolution is that we can use a given spacetime that is not a solution
of the Einstein equations for any reasonable matter, but that has some
properties worth investigating. In such a spacetime, we create a
fictitious stress-energy tensor simply defined by the Einstein
equations but not derived from any reasonable matter source.  We think
that such ``fake'' spacetimes can be valuable tests of qualitative
aspects of coordinate conditions.


\subsection{Stationary spacetimes}


We have used a variety of spacetimes to test three different gauges:
area gauge, generalized SY gauge and generalized BCT gauge.  The results
are as follows.

As model stationary spacetimes with a regular center we have examined
Minkowski space, de Sitter space and the Schwarzschild solution for a
constant density star, parameterized by $R_{\rm star}/M_{\rm
star}<9/4$.  All three gauge choices work on the Minkowski and de
Sitter spacetimes. The area and generalized SY gauges work on the
constant density star. The generalized BCT gauge fails numerically at
the boundary of the star, where $dg_{rr}/d r$ is discontinuous.  As a
boundary condition we choose $\alpha=1$ at the center. As the
equations are linear in both $\alpha$ and $\beta^r$, this choice is
equivalent to any other choice that is constant in time. As a
consistency check of our programming, we have verified that the
Hamiltonian and momentum constraints are obeyed up to finite
differencing error. As the initial slice, we take $X^1=r$ and $X^0(r)$
a Gaussian, with $X^0$ and $X^1$ the standard time and radial
coordinates. We find that the induced metric, extrinsic curvature,
stress-energy components, lapse and shift are all time-independent up
to numerical error.

To test coordinate conditions in a black hole spacetime with the black
hole excised from the numerical domain, we have implemented the
Schwarzschild spacetime in coordinates that are regular at the future
event horizon (the Painlev\'e-G\"ullstrand coordinates).  Here the
area gauge and generalized SY gauge work. The generalized BCT gauge
has numerical problems at the excision boundary. As a boundary
condition we have chosen $\alpha=1$ at the excision boundary, which is
chosen at constant coordinate $r$. As the shift is supposed to keep
our generic radial coordinate $r$ at constant curvature radius
$\sqrt{g_{\theta\theta}}$, this boundary condition is consistent.  As
the initial slice, we took $X^1=r$ and $X^0(r)$ a Gaussian, with $X^1$
the area radius and $X^0$ the Painlev\'e-G\"ullstrand time (which is
defined by surfaces of constant $X^0$ being flat and spacelike). Again
we find that all variables are constant in time, even though they are
all nontrivial.


\subsection{Self-similar spacetimes}


We have also investigated four self-similar spacetimes. Our coordinate
prescriptions are designed to bring the metric of a CSS spacetime into
the form
\begin{equation}
g_{\mu\nu} = e^{-2t} \bar g_{\mu\nu}
\end{equation}
where the components of $\bar g_{\mu\nu}$ are independent of $t$. The
metric of a DSS spacetime can be written in the same form, with $\bar
g_{\mu\nu}$ periodic in $t$. It is important to note that while all
our prescriptions must work on CSS spacetimes, barring numerical error
and numerical instability, it is not guaranteed that they work, in the
sense of bringing the metric into this form, on DSS spacetimes. The
reason is that given a CSS spacetime and an initial slice, the
coordinate system of this form is unique, while it is not for a DSS
spacetime. In order to see how successful a gauge is in making the
conformal spacetime metric independent of $t$ or periodice in $t$, it
is useful to rescale fields intrinsic to the slice ($g_{ij}$, $K_{ij}$
etc.) with appropriate powers of $\exp t$ so that the rescaled fields
are independent of time if and only if the spacetime is CSS and our
algorithm finds the conformal Killing vector.

The simplest CSS spacetime, Minkowski spacetime, is not suitable for
our investigation, as the homothetic vector is not unique.  Instead,
we have considered a conformally flat spacetime of the form
\begin{equation}
ds^2 = e^{-2t}\left(-dt^2+dr^2+r^2\,d\Omega^2\right).
\end{equation}
We have given this spacetime fictitious matter defined by the Einstein
equations. It has a unique homothetic vector field if by assumption we
restrict that vector field to the $tr$ plane. Here, we have tested all
three gauges. All work well when we use the the analytic boundary
condition $\alpha=e^{-t}$ at the center.

A less trivial example is the Roberts \cite{Roberts} solution, a
family of CSS massless scalar field solutions, which in double null
coordinates is given by
\begin{eqnarray}
ds^2 & = & -du\, dv + {1\over4}\left[(1-p^2)v^2 -2uv +
u^2\right]\,d\Omega^2 \\
\phi(u,v) & = & \ln {(1-p)v-u\over (1+p)v -u},
\end{eqnarray}
with $p$ a parameter.  This spacetime has curvature singularities at
$g_{\theta\theta}=0$. We have investigated the case $p=0.5$, where
these singularities are naked, and lie in the future and past
lightcones. The slice we evolve lies in the spacelike sector. The only
boundary condition for $\alpha$ at the inner boundary that worked here
is $\alpha=\alpha_0\exp\sigma t$, with $\alpha_0$ taken from the exact
solution. Here the area gauge and generalized SY gauge work. The
generalized BCT gauge has numerical problems at the excision boundary.

Our other two examples are of more physical interest. They are the CSS
critical solution for the collapse of a perfect fluid with the
equation of state $p=(1/3)\rho$ \cite{EvansColeman} and the DSS
critical solution for the collapse of a massless scalar field
\cite{Choptuik}. These solutions have been constructed numerically as
solutions of a nonlinear ODE eigenvalue problem \cite{EvansColeman},
and a nonlinear PDE eigenvalue problem \cite{critscalar},
respectively. We interpolate them numerically to the required values
of $X^0$ and $X^1$. For both spacetimes, we have tested the area gauge 
and the generalized SY gauge. Again, we use a Gaussian initial slice. 

In the CSS fluid spacetime, we have been able to use the boundary
condition (\ref{CSSalphacenter}). The boundary condition works, and all
variables are constant in time.

The DSS scalar field spacetime has a regular center, but the scalar
field is a periodic function of $\tau$ at the center, so that the
density $\rho$ vanishes twice per period. This means that we cannot
apply the boundary condition (\ref{CSSalphacenter}). Instead, we have
applied the exact boundary condition $\alpha=\alpha_0\exp\sigma t$,
with $\alpha_0$ taken from the exact solution. This test is
nontrivial, as it is not clear at all that a coordinate condition
designed to catch a continuous symmetry can catch a discrete one. But
it works: all variables are periodic in time for the scalar field DSS
spacetime for at least two periods, in both the area and generalized
SY gauge.


\section{Conclusions}


We have examined the problem of finding good coordinates in numerical
relativity, and the problem of defining a renormalisation group for
general relativity in critical collapse, in a single framework. We
have listed old and new prescriptions that obey our criterion for a
good coordinate condition in either application: that the presence of
an approximate timelike Killing field, or an approximate homothetic
vector field, in some region of the spacetime (for example at late
times) be revealed by metric coefficients that vary only slowly with
time.

We have tested several prescriptions numerically in spherically
symmetric spacetimes that are exactly stationary or homothetic. For
these exact symmetries, it is clear analytically that our
prescriptions will work, but we have used these cases both to test our
code, and to see how well different schemes can be implemented
numerically. We find that a scheme specific to spherical symmetry
(area freezing shift and radial lapse) is the most stable one,
generalized Smarr-York (SY) gauge (minimal strain shift and
exponential $K$ lapse) comes second, while we have been unable to
implement the generalized Brady-Crighton-Thorne (BCT) gauge (minimal
strain shift and conformal strain lapse) stably. What this means is
that the combination of our numerical methods of solving for the BCT
gauge and of evolving a slice is numerically unstable. This indicates
that coupling the BCT gauge to a genuine spacetime evolution code may
also be numerically unstable, but it is of course possible that a
different numerical implementation of the gauge and/or the evolution
equations is stable. Other difficulties of the BCT gauge are evident
analytically: Not all initial data sets (in our case, initial slices
through a given spacetime) admit a solution to BCT gauge, and BCT
gauge may still break down (when $K_{ab}K^{ab}=0$ anywhere) at a later
time. In spherical symmetry, the slice must nowhere be close to a
polar slice for BCT gauge to work.

We have entered new terrain examining one spacetime (the Choptuik
scalar field critical solution) with a discrete, rather than
continuous, symmetry. Here we found numerically that two prescriptions
made for a continuous symmetry (area gauge and generalized SY gauge)
actually follow the discrete symmetry, in the sense that the metric
becomes periodic in time. (Our third gauge, generalized BCT gauge, was
unstable numerically in our implementation, but may yet work with
better numerical methods.) This is an important step in the
construction of a renormalisation group for critical collapse, as it
suggests that a single coordinate prescription may realize the
renormalisation group both for continuously and discretely
self-similar (CSS and DSS) critical solutions. It is also encouraging
for the use of co-rotating coordinates in non-circular binary black
hole inspiral.

One crucial ingredient is still lacking in a working prescription for
a renormalisation group, namely the boundary conditions used when
solving elliptic equations for the lapse and shift. Details aside,
these boundary conditions must contain a good guess for how far away
in proper time the slice is from the accumulation point of the
(approximately) self-similar spacetime. Although there are ideas [see
equation (\ref{CSSalphacenter})], we have not been able to solve this
problem, and instead had to supply this information by hand.

Future work should include testing beyond spherical symmetry and in
particular with angular momentum, and on spacetimes with approximate
symmetries, as well as developing ideas on what boundary conditions
should be imposed on elliptic coordinate conditions.


\acknowledgements


We would like to thank Beverly Berger, Matt Choptuik, Bob Wald 
and Stu Shapiro for 
helpful discussions.  DG would like to thank the Albert Einstein
Institute and the University of Chicago for hospitality.  DG and CG
would like to thank the Institute for Theoretical Physics at Santa
Barbara (partially supported by NSF grant PHY-9407194) for hospitality.
This work was partially supported by NSF grant PHY-9514726 to the University
of Chicago and by NSF grant PHY-9722039 to Oakland University.



\end{document}